\documentstyle[psfig,12pt]{article}
\def\lsim{\:\raisebox{-0.5ex}{$\stackrel{\textstyle<}{\sim}$}\:}

\def\be{\begin{equation}}       
\def\ee{\end{equation}}
\def\bear{\be\begin{array}}      
\def\eear{\end{array}\ee}
\def\bea{\begin{eqnarray}}
\def\eea{\end{eqnarray}}

\include{mat}
\def\21{$SU(2) \ot U(1)$}
\def\ot{\otimes}
\def\ie{{\it i.e.}}
\def\etal{{\it et al.}}
\def\half{{\textstyle{1 \over 2}}}

\def\quarter{{\textstyle{1 \over 4}}}

\def\eighth{{\textstyle{1 \over 8}}}
\def\bold#1{\setbox0=\hbox{$#1$}
     \kern-.025em\copy0\kern-\wd0
     \kern.05em\copy0\kern-\wd0
     \kern-.025em\raise.0433em\box0 }
\begin{document}
\begin{titlepage}
\begin{flushright}
FTUV/97-56\\
IFIC/97-72\\
hep-ph/9710233\\
September 1997
\end{flushright}
\vspace*{5mm}
\begin{center} 
{\Large \bf Charged Higgs Sector with and without R--Parity}\\[15mm]
{\large Marco Aurelio D\'\i az} \\
\hspace{3cm}\\
{\small Departamento de F\'\i sica Te\'orica, IFIC-CSIC, Universidad de Valencia}\\ 
{\small Burjassot, Valencia 46100, Spain}
\end{center}
\vspace{5mm}
\begin{abstract}

The simplest way of studying systematically R--parity violating phenomena
is by introducing a bilinear term in the superpotential of the type
$\epsilon\widehat L\widehat H_2$, which violates R--parity and lepton 
number but keep barion number conserved. In its simplest version, this 
``$\epsilon$--model'' is a two parameter extension of the MSSM and a
one parameter extension of the MSSM--SUGRA. Here we study the charged Higgs
sector of the model, which mixes with the stau sector, and compare it with
the charged Higgs sector of the MSSM. We demonstrate that $m_{H^{\pm}}$
can be lower than $m_W$ already at tree level, and calculate the production
cross section of pairs of charged Higgs and staus. In this model it is 
possible the mixed production of a charged Higgs and a stau and its
production cross section can be sizable. We finally comment about the
new R--parity violating decay modes of the charged scalars.

\end{abstract}

\vskip 5.cm
\noindent ${}^{\dag}$ Talk given at the Third Warsaw Workshop ``Physics
from Planck Scale to Electroweak Scale'', Warsaw University, 2-5 April
1997, Warsaw, Poland.

\end{titlepage}

\setcounter{page}{1}

\section{Introduction}

After the discovery of the top quark at Fermilab \cite{topDisc}, the
Higgs boson is the only particle predicted by the Standard Model (SM)
still undetected. Theory does not predict the mass $m_H$ of the Higgs
boson, but it should satisfy $m_H\lsim 1$ TeV, otherwise the Higgs
sector becomes non--perturbative. On the other hand, an experimental lower
bound of $m_H>77$ GeV has been set by LEP \cite{LEPmhSM} due to its
non--observation in $e^+e^-$ collisions. Despite its simplicity, 
the SM Higgs sector is not attractive theoretically because the Higgs 
boson mass is unstable under radiative corrections. The most popular
extension of the SM that solves this problem is the Minimal Supersymmetric
Standard Model (MSSM), whose Higgs sector contains two CP--even neutral
Higgs bosons $h$ and $H$, a CP--odd Higgs boson $A$, and a pair of charged 
Higgs bosons $H^{\pm}$.

Here we are interested in the phenomenology of the charged Higgs boson
\cite{ChaStau} in a model which violates R--parity through a bilinear term 
of the form $\varepsilon_{ab}\epsilon_i\widehat L_i^a\widehat H_2^b$ in the 
superpotential \cite{HallSuzuki}. The superpotential we consider is
\begin{equation} 
W=\varepsilon_{ab}\left[
 h_U^{ij}\widehat Q_i^a\widehat U_j\widehat H_2^b
+h_D^{ij}\widehat Q_i^b\widehat D_j\widehat H_1^a
+h_E^{ij}\widehat L_i^b\widehat R_j\widehat H_1^a
-\mu\widehat H_1^a\widehat H_2^b
+\epsilon_i\widehat L_i^a\widehat H_2^b\right]
\label{eq:Wsuppot}
\end{equation}
where the first four terms correspond to the MSSM and the last one is
the bilinear term which violates R--parity \cite{e3others}. From now on, 
the model described by the superpotential in eq.~(\ref{eq:Wsuppot}) will 
be called here the $\epsilon$--model. For simplicity we consider the case 
where $\epsilon_1=\epsilon_2=0$ and $\epsilon_3\ne 0$. This model is a 
truncated version of a more complete model in which a vacuum expectation
value of a right--handed sneutrino field induces the $\epsilon$--term
\cite{ComVersion}.

In the $\epsilon$--model, a vacuum expectation value $v_3$ of the 
left--handed sneutrino--tau field is induced by the $\epsilon$--term and, 
therefore, the lepton number is not conserved. In addition,
the tau--neutrino mixes with the neutralinos and the $\nu_{\tau}$ acquires
a mass which satisfy $m_{\nu_{\tau}}\sim (\mu v_3+\epsilon_3 v_1)^2$,
where $v_1$ is the vev of $H_1$ and $v_3$ is the vev of $\tilde\nu_{\tau}$. 
We need a small value of the neutrino--tau 
mass in order to satisfy the experimental upper bound on $m_{\nu_{\tau}}$,
and this is achieved without a fine tunning on the combination
$(\mu v_3+\epsilon_3 v_1)$. The reason is that with universality of 
scalar masses and universality of bilinear soft terms, the smallness
of the combination $(\mu v_3+\epsilon_3 v_1)$ is natural and, in fact, 
proportional to $h_b^2/8\pi^2$, where $h_b$ is the bottom quark Yukawa
coupling. When embedded into supergravity, this $\epsilon$--model is a 
one parameter extension of the MSSM--SUGRA and, therefore, the simplest
way to study systematically R--parity violating phenomena \cite{epsrad}.

The $\epsilon$--term in eq.~(\ref{eq:Wsuppot}) cannot be rotated away by 
a redefinition of the superfields $\widehat H_1'=
(\mu\widehat H_1-\epsilon_3\widehat L_3)/\sqrt{\mu^2+\epsilon_3^2}$.
If this rotation is performed, the bilinear term disappears from the
superpotential, but trilinear R--parity violating terms are re--introduced.
Furthermore, if supersymmetry is broken, bilinear terms which induce a
vacuum expectation value of the sneutrino--tau are re--introduced in the 
scalar sector. This occurs even if universality of scalar masses and 
universality of bilinear soft breaking terms hold at the unification scale.

Also important in the $\epsilon$--model is the fact that the charginos 
mix with the tau lepton, the CP--even Higgs bosons mix with the real 
part of the $\tilde\nu_{\tau}$ field, the CP--odd Higgs bosons mix with
the imaginary part of $\tilde\nu_{\tau}$, and the charged Higgs bosons 
mix with the staus. In this note we analize some aspects of the charged
Higgs sector, first within the MSSM, and later in the $\epsilon$--model.

\section{Charged Higgs in the MSSM}

\begin{figure}
\centerline{\protect\hbox{\psfig{file=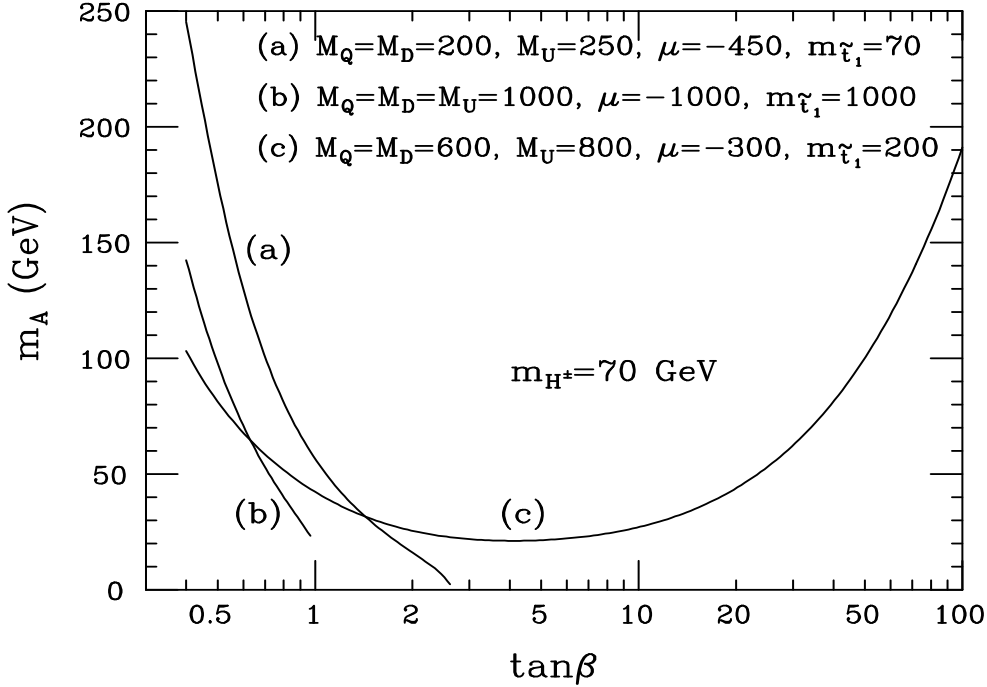,height=11cm,width=1.0\textwidth,angle=90}}}
\caption{
Radiatively corrected CP--odd mass $m_A$ as a function of $\tan\beta$ for
a constant value of the charged Higgs mass $m_{H^{\pm}}=70$ GeV in the 
MSSM.} 
\label{fig:MA_tb}
\end{figure} 
The MSSM charged Higgs boson has a one--loop corrected mass given 
by \cite{DiazHaberi,chhothers}
\begin{equation}
m_{H^{\pm}}^2=m_W^2+m_A^2+{\mathrm{Re}}\Big[A_{H^+H^-}(m_{H^{\pm}}^2)-
A_{WW}(m_W^2)-A_{AA}(m_A^2)\Big]\,,
\label{eq:mcha}
\end{equation}
where the charged Higgs mass $m_{H^{\pm}}$ is the solution to an implicit
equation. The tree level approximation is adequate in a large region of 
parameter space, specially if $m_A\gg m_Z$. Nevertheless, there are regions 
where one--loop contributions are crucial. Furthermore, it is possible that 
the charged Higgs mass is smaller than the tree level lower bound given by 
$m_{H^{\pm}}>m_W$. 
As an example, we consider the case in 
Fig.~\ref{fig:MA_tb}. In this case we take as an input the charged Higgs
mass $m_{H^{\pm}}=70$ GeV and calculate the one--loop corrected CP--odd
Higgs mass $m_A$. This is done simply by solving the implicit equation for 
$m_A$ given in
eq.~(\ref{eq:mcha}). Note that this value of the charged Higgs mass is
not possible at the classical level, because in that case $(m_A^2)_{tree}$
is negative. But quantum corrections are large enough to lift $m_A^2$ to
positive values, as we can see for three different choices of the
parameters that control the squark sector. Two of the curves are truncated
after taking into account the constraints from color breaking minima
\cite{colorBM}.

\begin{figure}
\centerline{\protect\hbox{\psfig{file=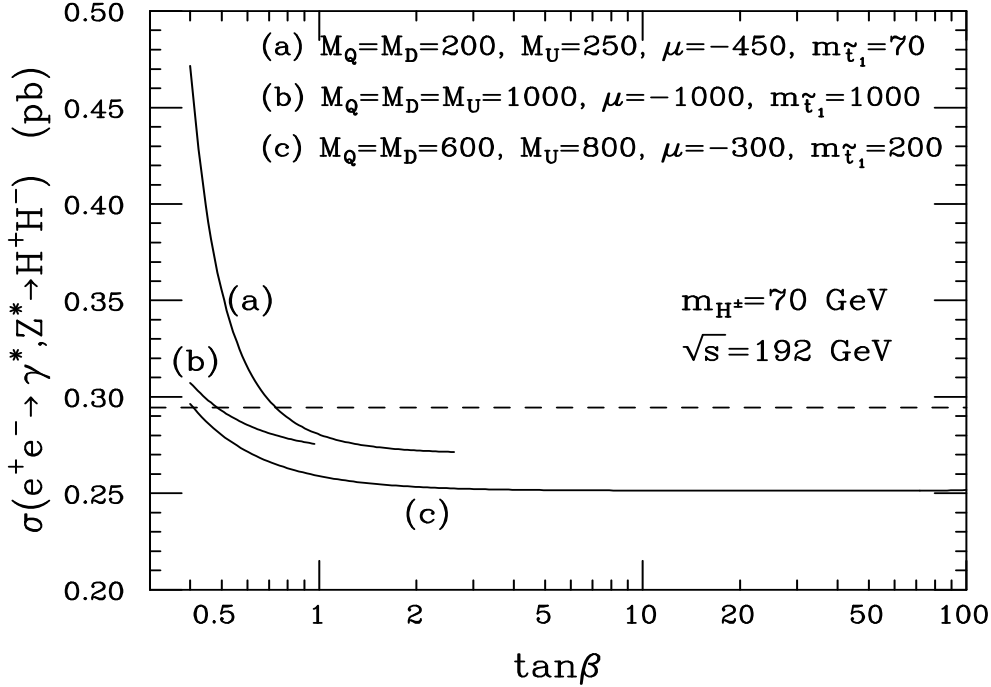,height=11cm,width=1.0\textwidth,angle=90}}}
\caption{
Tree level (dashes) and radiatively corrected (solid) total production cross
section of a pair of charged Higgs bosons at LEP2 as a function of 
$\tan\beta$ in the MSSM.} 
\label{fig:Sch_tb}
\end{figure} 
In the MSSM the charged Higgs pair production cross section depends only
on the charged Higgs mass. Nevertheless, radiative corrections can be 
important \cite{SigChaRC,DiazTonnis}. For the same choice of parameters in 
the previous figure, we have plotted the one--loop corrected cross section 
$\sigma(e^+e^-\rightarrow H^+H^-)$ in Fig~\ref{fig:Sch_tb}. Corrections
from $-15\%$ up to $+50\%$ can be appreciated, and may have an impact
on charged Higgs searches at LEP2. 

Charged Higgs decay modes in the MSSM are well known. If $\tan\beta$ is
small $H^+\rightarrow c\overline{s}$ dominates, and if $\tan\beta$ is
large $H^+\rightarrow \tau^+\nu_{\tau}$ is the dominant decay mode.
The decay mode $H^+\rightarrow c\overline{b}$ should also be considered.
QCD corrections to the decay width can be implemented with the running 
quark masses \cite{MenPoma}, and electroweak radiative corrections are 
small \cite{RCchadec}.

\section{Charged Higgs and Staus in the $\epsilon$--Model}

In the presence of a bilinear R--parity violating term in the superpotential,
as given in eq.~(\ref{eq:Wsuppot}), the phenomenology of the charged Higgs
sector changes dramatically. First of all, the charged Higgs boson fields
mix with the stau sector. The mass terms in the lagrangian are
\begin{equation}
V_{quadratic}=[H_1^-,H_2^-,\tilde\tau_L^-,\tilde\tau_R^-]
\bold{M_{S^{\pm}}^2}\left[\matrix{H_1^+ \cr H_2^+ \cr \tilde\tau_L^+ \cr
\tilde\tau_R^+}\right]+...
\label{eq:Vquadratic}
\end{equation}
where, for simplicity we decompose the charged scalar mass matrix into
three terms
\begin{equation}
\bold{M_{S^{\pm}}^2}={\bold M_{H^{\pm}}^2}+{\bold M_{{\tau}^{\pm}}^2}+
{\bold M_{\epsilon_3}^2}\,.
\label{eq:SmassDecom}
\end{equation}
In the first matrix ${\bold M_{H^{\pm}}^2}$ we include the MSSM terms 
relevant to the charged Higgs sector, and it is given by 
($g_Z^2\equiv g^2+g'^2$):
\begin{equation}
{\bold M^2_{H^{\pm}}}=\left[\matrix{
m_{H_1}^2+\mu^2+\quarter g^2v_2^2+\eighth g_Z^2(v_1^2-v_2^2) & 
B\mu+\quarter g^2v_1v_2 & 0 & 0 \cr 
B\mu+\quarter g^2v_1v_2 & 
m_{H_2}^2+\mu^2+\quarter g^2v_1^2-\eighth g_Z^2(v_1^2-v_2^2) & 0 & 0 \cr
0 & 0 & 0 & 0 \cr
0 & 0 & 0 & 0
}\right]
\label{eq:MchaMSSM}
\end{equation}
Similarly, we include in ${\bold M_{{\tau}^{\pm}}^2}$ the MSSM terms relevent
for the stau sector, and they are
\begin{equation}
{\bold M^2_{\tau^{\pm}}}=\left[\matrix{
0 & 0 & 0 & 0 \cr 
0 & 0 & 0 & 0 \cr
0 & 0 & 
m_{L_3}^2+\half h_{\tau}^2v_1^2-\eighth(g^2-g'^2)(v_1^2-v_2^2) & 
{1\over{\sqrt{2}}}h_{\tau}(A_{\tau}v_1-\mu v_2) \cr
0 & 0 & 
{1\over{\sqrt{2}}}h_{\tau}(A_{\tau}v_1-\mu v_2) & 
m_{R_3}^2+\half h_{\tau}^2v_1^2-\quarter g'^2(v_1^2-v_2^2)
}\right]
\label{eq:MstauMSSM}
\end{equation}
Finally, the terms not present in the MSSM and induced by the 
$\epsilon_3$--term in the superpotential are given by
\begin{eqnarray}
&& {\bold M^2_{\epsilon_3}}=
\label{eq:MepsMSSM} \\ \nonumber \\
&& \!\!\!\!\!\!\!\!\!\!\!\!\!\!\left[\matrix{
\half h_{\tau}^2v_3^2-\eighth(g^2-g'^2)v_3^2 & 0 & 
\quarter(g^2-2h_{\tau}^2)v_1v_3-\mu\epsilon_3 & 
-{1\over{\sqrt{2}}}h_{\tau}(\epsilon_3v_2+A_{\tau}v_3) \cr 
0 & \epsilon_3^2+\eighth(g^2-g'^2)v_3^2 & 
-B_2\epsilon_3+\quarter g^2v_2v_3 & 
-{1\over{\sqrt{2}}}h_{\tau}(\mu v_3+\epsilon_3v_1) \cr
\quarter(g^2-2h_{\tau}^2)v_1v_3-\mu\epsilon_3 & 
-B_2\epsilon_3+\quarter g^2v_2v_3 & 
\epsilon_3^2+\eighth(g^2+g'^2)v_3^2 & 0 \cr
-{1\over{\sqrt{2}}}h_{\tau}(\epsilon_3v_2+A_{\tau}v_3) & 
-{1\over{\sqrt{2}}}h_{\tau}(\mu v_3+\epsilon_3v_1) & 0 & 
\half h_{\tau}^2v_3^2-\quarter g'^2v_3^2
}\right] \nonumber \\ \nonumber
\end{eqnarray}
which is identical to zero in the MSSM limit where $\epsilon_3=v_3=0$. 
The charged scalar mass matrix $\bold{M_{S^{\pm}}^2}$ in 
eq.~(\ref{eq:SmassDecom}) is diagonalized by a rotation defined
by $\mathrm{diag}(0,m_{H^{\pm}}^2,m_{\tilde\tau_1^{\pm}}^2,
m_{\tilde\tau_2^{\pm}}^2)={\bold R_{S^{\pm}}}{\bold M_{S^{\pm}}^2}
{\bold R_{S^{\pm}}^T}$, and the mass eigenstates are
\begin{equation}
\left[\matrix{S^+_1 \cr S^+_2 \cr S^+_3 \cr S^+_4}\right]\equiv
\left[\matrix{G^+ \cr H^+ \cr \tilde\tau_1^+ \cr \tilde\tau_2^+}\right]=
{\bold R_{S^{\pm}}}\left[\matrix{
H_1^+ \cr H_2^+ \cr \tilde\tau_L^+ \cr \tilde\tau_R^+}\right]
\label{eq:eigenvectors}
\end{equation}
where $G^+$ is the massless Goldstone boson, $\tilde\tau^+_1$ and 
$\tilde\tau^+_2$ are the two mass eigenstates with the largest stau 
component, and the remaining mass eigenstate is the charged Higgs $H^+$.
The determinant of the matrix $\bold{M_{S^{\pm}}^2}$ is explicitely zero
only after imposing the minimization conditions, or tadpole equations, which
in the $\epsilon$--model are equal to
\begin{eqnarray}
t_1^0&=&(m_{H_1}^2+\mu^2)v_1-B\mu v_2-\mu\epsilon_3v_3+
\eighth(g^2+g'^2)v_1(v_1^2-v_2^2+v_3^2)\,,
\nonumber \\
t_2^0&=&(m_{H_2}^2+\mu^2+\epsilon_3^2)v_2-B\mu v_1+B_2\epsilon_3v_3-
\eighth(g^2+g'^2)v_2(v_1^2-v_2^2+v_3^2)\,,
\label{eq:tadpoles} \\
t_3^0&=&(m_{L_3}^2+\epsilon_3^2)v_3-\mu\epsilon_3v_1+B_2\epsilon_3v_2+
\eighth(g^2+g'^2)v_3(v_1^2-v_2^2+v_3^2)\,.
\nonumber
\end{eqnarray}
at tree level. At the minimum we must impose $t^0_i=0$, $i=1,2,3$.

\begin{figure}
\centerline{\protect\hbox{\psfig{file=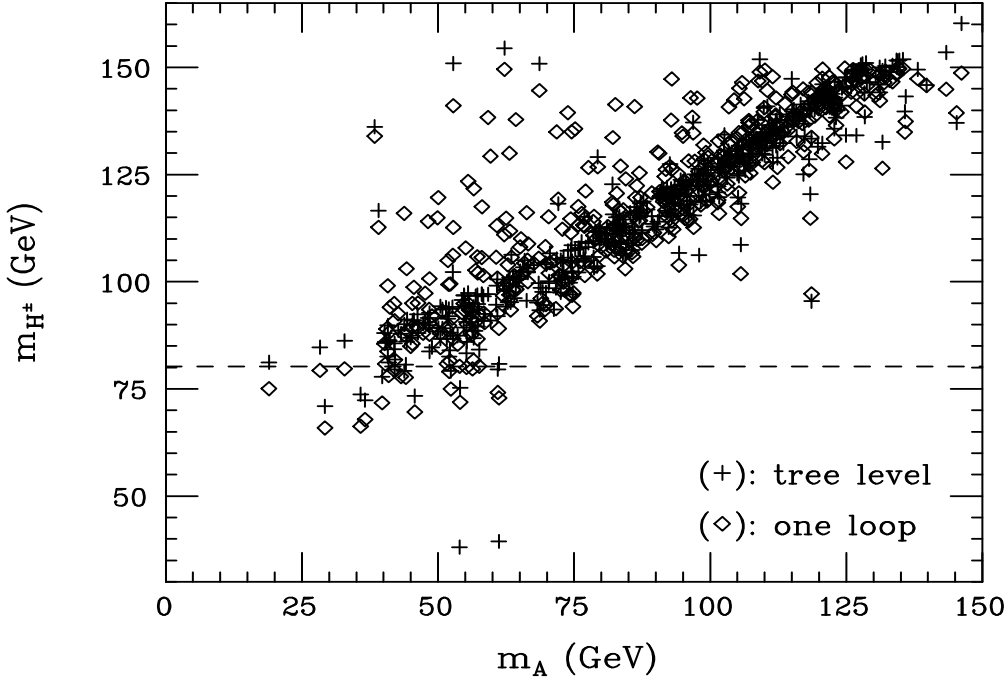,height=11cm,width=1.0\textwidth,angle=90}}}
\caption{
Tree level and one--loop charged Higgs mass as a function of $m_A$ in the 
$\epsilon_3$--model.} 
\label{fig:v3mchma}
\end{figure} 
We calculate the charged Higgs mass $m_{H^{\pm}}$ and we find that already at
tree level this mass can be lighter that $m_W$, as opposed to the MSSM case.
This can be understood if we work in the approximation 
$|v_3|\ll |\epsilon_3|\ll m_W$ and $h_{\tau}=0$, where we find
\begin{equation}
m_{H^{\pm}}^2-m_A^2\approx m_W^2+(\mu^2+B_2^2)\left[
{1\over{m_{H^{0\pm}}^2-m_{\tilde\tau_L^0}^2}}-
{1\over{m_{A^0}^2-m_{\tilde\nu_{\tau}^0}^2}}\right]\epsilon_3^2
\label{eq:mhmae3}
\end{equation}
Here the masses $m_{H^{0\pm}}$, $m_{\tilde\tau_L^0}$, $m_{A^0}$, and
$m_{\tilde\nu_{\tau}^0}$ are calculated in the limit $v_3=\epsilon_3=0$,
\ie, in the MSSM. Using standard relations it can be shown that
$m_{H^{\pm}}^2-m_A^2<m_W^2$ already at tree level. This can be appreciated
in Fig.~\ref{fig:v3mchma} where we make a scan of $10^3$ points in parameter
space: some of the tree level points are below the dashed
line corresponding to the MSSM lower limit $m_{H^{\pm}}=m_W$. We impose that
the induced tau--neutrino mass is smaller than 30 MeV.

In the same Fig.~\ref{fig:v3mchma} we plot the one--loop renormalized
charged Higgs mass. To calculate it, we work in the MSSM approximation
where we add to the tree level mass the correction given in 
eq.~(\ref{eq:mcha}). The number of points below the MSSM lower limit
increases after adding radiative corrections because these corrections 
are negative if $\tan\beta$ is small. There are also many points where 
quantum corrections are large and positive. This happens when $\tan\beta$ 
is large.

\begin{figure}
\centerline{\protect\hbox{\psfig{file=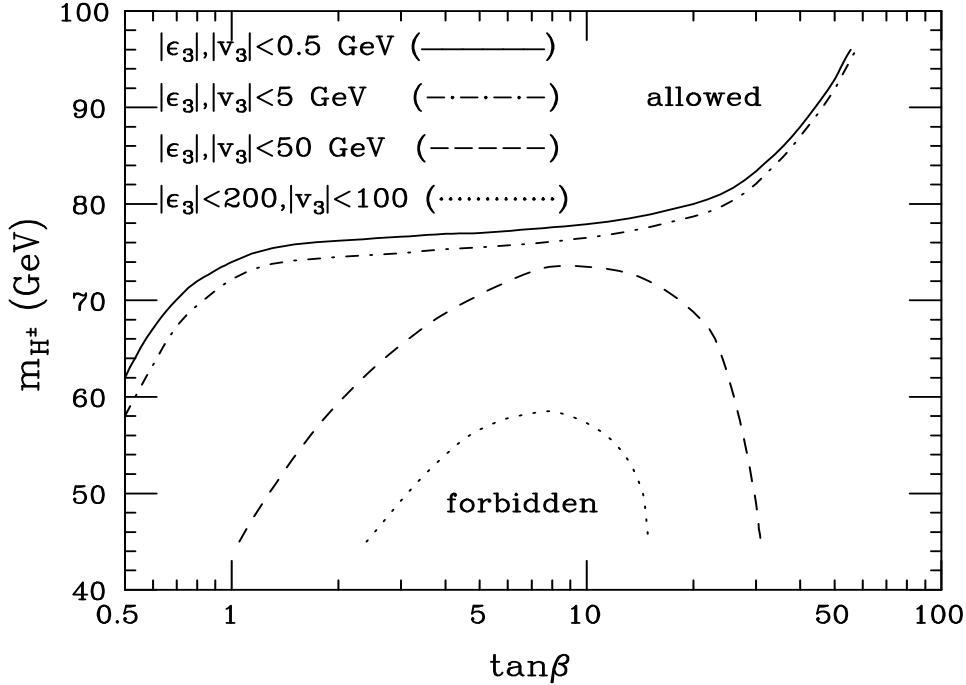,height=11cm,width=1.0\textwidth,angle=90}}}
\caption{
Possible values of the one--loop corrected charged Higgs mass as a function
of $\tan\beta$ for different upper bound for $\epsilon_3$ and $v_3$.} 
\label{fig:v3mchtb}
\end{figure}

The effect that the $\epsilon_3$--term has on the charged Higgs mass can
be appreciated also in Fig.~\ref{fig:v3mchtb}. Here we make a scan over 
parameter space with $10^4$ points, and the curves are the boundary below
which no points are found (``forbidden'' region). In solid line we take
$\epsilon_3<0.5$ and $v_3<0.5$ GeV and the situation is pretty close
to the MSSM case. The minimum value of the charged Higgs mass is around
$m_W$, but radiative corrections decrease this minimum value when 
$\tan\beta$ is small, and increase it if $\tan\beta$ is very large.
As soon as the values of $\epsilon_3$ and $v_3$ are increased, the charged
Higgs mass can take lower and lower values, as indicated by the other
curves. We stress the fact that the ``forbidden'' regions are in reality
regions where the scan did not find any solution.

\begin{figure}
\centerline{\protect\hbox{\psfig{file=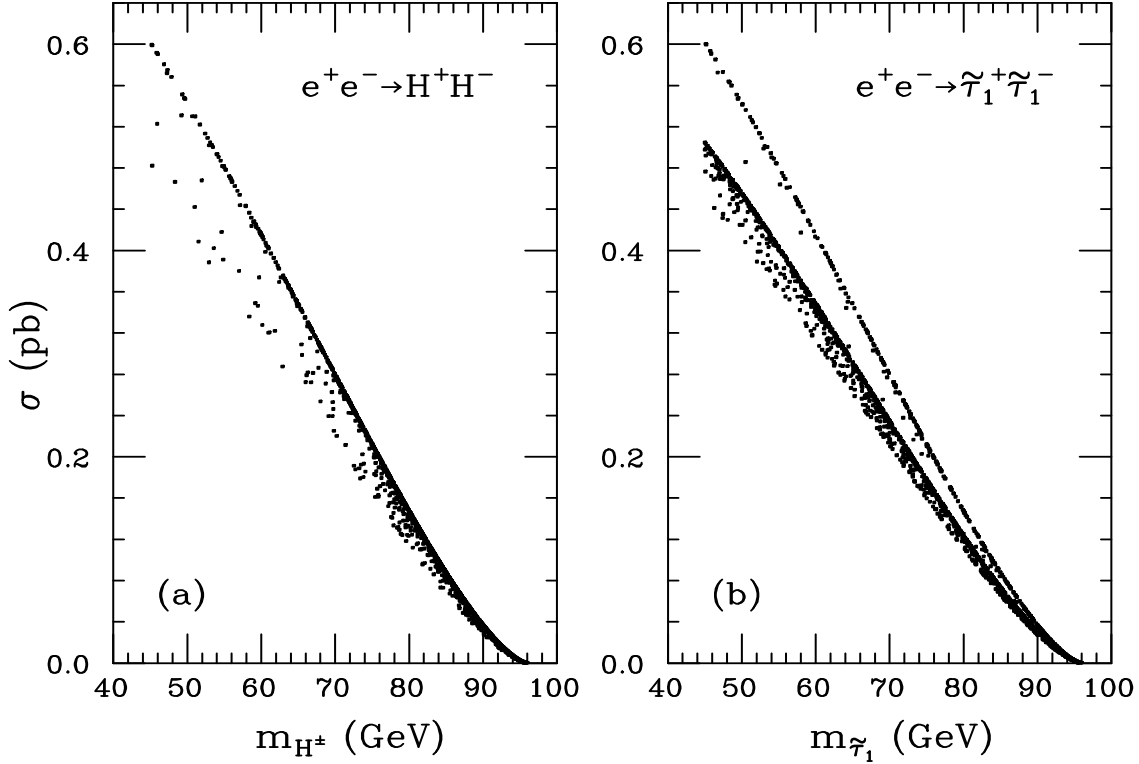,height=11cm,width=1.0\textwidth,angle=90}}}
\caption{
Total production cross section of a pair of (a) charged Higgs and (b) light
staus, as a function of their mass.} 
\label{fig:v3Spair}
\end{figure}

The production cross sections of pairs of charged scalars are also modified
by the $\epsilon_3$ term. In Fig.~\ref{fig:v3Spair}a we plot the charged
Higgs pair production cross section $\sigma(e^+e^-\rightarrow H^+H^-)$
as a function of the charged Higgs mass $m_{H^{\pm}}$, made with a scan of
$4\times 10^4$ points in parameter space. Most of the points fall into
the MSSM curve, but there are some deviations due to charged Higgs fields
mixing with right--stau. This can be easily understood if we look at the
$Z$ coupling to a pair of charged scalars. The $ZS_i^+S_j^-$ Feynman rule 
is equal to $i\lambda_{ZS^+S^-}^{ij}(p+p')^{\mu}$ where $p$ and 
$p'$ are the momenta in the direction of the positive electric charge 
flow. The $\lambda$ couplings are equal to ${\bold\lambda}_{ZS^+S^-}={\bold 
R_{S^{\pm}}} {\bold\lambda'_{ZS^+S^-}}{\bold R_{S^{\pm}}^T}$ where the
couplings in the unrotated basis are 
\begin{equation}
{\bold\lambda'_{ZS^+S^-}}={g\over{2c_W}}\left[\matrix{ -c_{2W} & 0 & 0
& 0 \cr 0 & -c_{2W} & 0 & 0 \cr 0 & 0 & -c_{2W} & 0 \cr 0 & 0 & 0 &
2s_W^2 }\right] \label{eq:ZSScouplings} 
\end{equation} 
If the right stau were decoupled from the rest of the charged
scalars, the charged Higgs pair production cross section would be 
identical to the MSSM for any value of $\epsilon_3$ or $v_3$. The 
reason is that the upper--left $3\times 3$ relevant sub-matrix of 
$\bold\lambda'_{ZS^+S^-}$ in eq.~(\ref{eq:ZSScouplings}) is proportional 
to the identity.

In Fig.~\ref{fig:v3Spair}b we plot the total production cross section
of a pair of light staus $\tilde\tau^{\pm}_1$. The points concentrate 
around the two MSSM curves corresponding to the production of left--staus
(upper curve) and right--staus (lower curve). The cross sections, calculated
for a center of mass $\sqrt{s}=192$ GeV, have a maximum value of 0.6 pb
for any of the charged scalars.

\begin{figure}
\centerline{\protect\hbox{\psfig{file=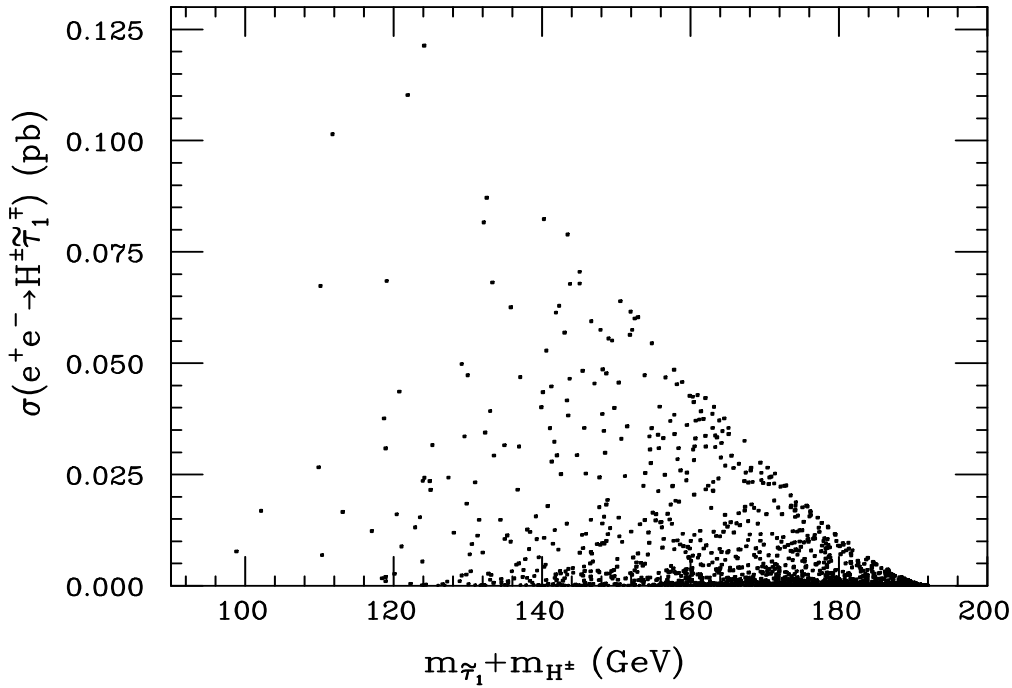,height=11cm,width=1.0\textwidth,angle=90}}}
\caption{
Total mixed production cross section of a charged Higgs and a light stau
as a function of the sum of their masses.} 
\label{fig:v3Smix}
\end{figure}

It is very interesting to notice that, contrary to what happens in the MSSM,
the mixed production of a charged Higgs and a stau is allowed in the 
$\epsilon$--model. In Fig.~\ref{fig:v3Smix} we plot the mixed cross
section $\sigma(e^+e^-\longrightarrow H^{\pm}\tilde\tau_1^{\mp})\equiv
\sigma(e^+e^-\longrightarrow H^+\tilde\tau_1^-)+
\sigma(e^+e^-\longrightarrow H^-\tilde\tau_1^+)$ as a function of the total
mass of the products $m_{H^{\pm}}+m_{\tilde\tau^{\pm}_1}$ for a center
of mass energy $\sqrt{s}=192$ GeV.
The mixed cross section has a sizable maximum value of 0.125 pb, and as \
espected it is smaller than the maximum value of the pair production cross 
section.

Decay modes of charged scalar particles are modified with respect to
the MSSM for two reasons. First, there are new R--parity violating
channels like $H^+\rightarrow \tilde\chi^0_1 \tau^+$ and
$H^+\rightarrow \tilde\chi^+_1 \nu_{\tau}$ for the charged Higgs,
and $\tilde\tau^+_i\rightarrow \nu_{\tau} \tau^+$ and
$\tilde\tau^+_i\rightarrow c\overline{s}$ for the staus. And second,
the lightest supersymmetric particle is not stable. In the case of
the neutralino as the LSP, its decay modes are 
$\tilde\chi^0_1\rightarrow \nu_{\tau}Z^*\rightarrow 
\nu_{\tau}q\overline{q}(l\overline{l})$ and 
$\tilde\chi^0_1\rightarrow \tau W^*\rightarrow 
\tau q\overline{q'}(l\overline{\nu_l})$. Furthermore, the LSP need not 
to be the lightest neutralino, and if the LSP is the lightest stau, it
can have R--parity violating decays 
$\tilde\tau^+_1\rightarrow \nu_{\tau} \tau^+(c\overline{s})$ with
a $100\%$ branching ratio \cite{ChaStau}.

\section{Conclusions}

In the ``$\epsilon$--model'', where R--parity and lepton number are
violated by the introduction of a bilinear term of the type
$\varepsilon_{ab}\epsilon_i\widehat L_i^a\widehat H_2^b$ in the
superpotential, the charged Higgs sector mixes with the stau sector.
The mass eigenstates are a set of four charged scalars: the unphysical
massless Goldstone boson $G^{\pm}$, the charged Higgs boson $H^{\pm}$,
and two staus $\tilde\tau^{\pm}_i$, $i=1,2$. We showed that the
charged Higgs mass $m_{H^{\pm}}$ can be lower than $m_W$ even at tree
level, contrary to what happens in the MSSM. Including radiative corrections,
the charged Higgs mass can be as light as 45 GeV. Values of $\epsilon_3$
and $v_3$ of the order of 100 GeV are compatible with the induced 
tau--neutrino mass. New R--parity violating decay modes are allowed,
where the charged Higgs can decay into charginos or neutralinos and 
staus can decay into quarks or leptons. The lightest supersymmetric particle
need not to be $\tilde\chi^0_1$ because it is not stable. If 
$\tilde\tau^{\pm}_1$ is the LSP, then it can decay into the R--parity
violating decay modes 
$\tilde\tau^+_1\rightarrow \nu_{\tau} \tau^+(c\overline{s})$
with $100\%$ branching fraction, no matter how small the R--parity
violating parameters $\epsilon_3$ and $v_3$ are. We claim that the
``$\epsilon$--model'' is the simplest way of studying systematically
R--parity violating phenomena, and in its simplest version, it is a
two parameter extension of the MSSM and a one parameter extension of the 
MSSM--SUGRA.

\section*{Acknowledgements}

I am thankful to Howard E. Haber and Tonnis ter Veldhuis for their
contribution to the R--parity conserving section of this work, and to
A. Akeroyd, J. Ferrandis, M. A. Garc{\'\i}a-Jare\~no, and Jos\'e W. F. 
Valle for their contribution to the R--parity violating section.
This work was supported by a DGICYT postdoctoral grant and by the
TMR network grant ERBFMRXCT960090 of the European Union.

\end{document}